\documentclass[aps,pre,preprint,superscriptaddress]{revtex4-1}  
\usepackage{graphicx}
\usepackage{subcaption}
\usepackage{dcolumn}
\usepackage{amssymb}
\usepackage{amsmath}
\usepackage{newtxtext,newtxmath,lipsum}
\usepackage[shortlabels]{enumitem}

\begin{document}
	\title{Breather and Positon excitations in a nonlinear electrical transmission line modeled by the Kundu-Eckhaus Equation}
	
	\author{N. Sinthuja}
	\affiliation{Department of Physics, Anna University, Chennai - 600 025, Tamilnadu, India}
	
	\author{M. Senthilvelan}
	\affiliation{Department of Nonlinear Dynamics, Bharathidasan University, Tiruchirappalli - 620 024, Tamilnadu, India}

    \author{K. Murali}
	\affiliation{Department of Physics, Anna University, Chennai - 600 025, Tamilnadu, India}
    
	
	\begin{abstract}
		\par  In this study, we explore the dynamics of breathers and positons in a nonlinear electrical transmission line modeled by the modified Naguchi circuit, governed by the Kundu-Eckhaus equation. Utilizing the reductive perturbation method and a specific transformation, we analyze the influence of different time-dependent linear potentials on these nonlinear wave structures. The analysis is conducted for three representative cases: (i) a constant potential, which modifies the orientation and amplitude of breathers and positons, (ii) a periodically modulated potential, which transforms them into crescent-shaped structures with unique spatial characteristics, and (iii) an exponentially varying potential, which induces asymmetric crescent-shaped waveforms. Additionally, we show that linear potentials significantly influence breather and positon dynamics in the modified electrical transmission line by altering their position and positon amplitude—constant potentials maintain peaks at the origin, periodic potentials shift breathers forward and positons backward, while exponential potentials move breathers backward and positons forward. Our findings highlight the critical role of external modulation in shaping wave propagation, localizing waves, and altering their amplitude, demonstrating its potential for controlling wave dynamics in nonlinear transmission lines. Unlike previous studies that focused on rogue waves, this work provides new insights into the evolution of breathers and positons under external perturbations. The results may have significant implications for applications in electrical transmission networks.
	\end{abstract}
	
	\maketitle

\section{Introduction}\label{sec1}
Nonlinear wave dynamics play a fundamental role in various physical systems that arise in fluid dynamics, and electrical circuits \cite{Yang2010nonlinear,Solli2007optical,Dudley2019rogue,Luo2020solitons,Ricketts2018electrical}. Among the well-known nonlinear wave structures, breathers and positons emerge as intriguing solutions of the nonlinear Schr\"odinger (NLS) equation, governing localized and periodic wave modulations \cite{Beutler1994what,Thulasidharan2024examining,Rasinariu1996negaton,Solution1,Solution2}. These solutions have been extensively studied in optics and oceanography, serving to model mechanisms such as modulational instability and the emergence of extreme waves \cite{Qiu2019the,Guo2019darboux,Matveev1992positon,Dubard2010on,Stahlhofen1995positons,Beautler1993positon,Song2019generating,Zhang2020novel}. However, despite the extensive research on soliton and rogue wave formation in nonlinear systems, the study of breathers and positons remains limited in the context of electrical transmission lines \cite{3,Aziz2020analytical,Kengne2022ginzeburg,Pelap2015dynamics,Marquie1994generation,Djelah2023first,Kengne2017modelling,Kengne2019transmission,Guy2018construction}.

Breather solutions have been previously explored in a few works related to nonlinear electrical circuits \cite{Guy2018construction}, but positon solutions, which exhibit a fundamentally different evolution characterized by growing oscillatory structures, have not yet been considered in this setting. Recently, we investigated the transmission of breathers and super rogue waves in a modified Noguchi electrical transmission line modeled by the NLS equation. We further extended our analysis to positon transmission, revealing interesting nonlinear wave interactions in this circuit system. However, the standard NLS equation provides a limited framework due to the absence of higher-order nonlinear terms and free parameters, which restrict the control over wave dynamics and the diversity of accessible solutions.

To address this limitation, we now model the modified Noguchi electrical transmission line as the Kundu-Eckhaus (KE) equation, an integrable generalization of the NLS equation. The KE equation incorporates both quintic nonlinearity and a higher-order dispersion introducing additional physical effects that influence wave propagation \cite{Kundu1, Kundu2, Kundu3}. This enriched mathematical structure enables the exploration of a broader class of nonlinear excitations, including breathers and positons, which remain largely unexplored in the context of electrical transmission lines. By studying these solutions in the KE framework, we aim to uncover new dynamical behaviors that are inaccessible within the standard NLS model. Our results provide new insights into the role of higher-order nonlinear effects in electrical transmission lines, contributing to the broader understanding of nonlinear wave dynamics in engineered systems. 

To analyze these effects, we derive exact expressions for breather and positon waves for the KE and investigate their transmission properties in the nonlinear electrical circuit. Our results provide new insights into the role of higher-order nonlinear effects in electrical transmission lines, with potential implications for signal propagation and energy transport in nonlinear electrical networks.

The subsequent sections of this paper are organized as follows:  In Sec. II, we present the theoretical framework and derive the breather and positon solutions. In Sec. III, we analyze the effects of external potentials such as constant, periodic, and exponential on the breather and positon solutions. In Sec. IV, we study the dynamics of breather and positon solutions with external linear potentials in the modified electrical transmission line.  Finally, in Sec. V, we discuss the implications of our results and provide concluding remarks.

\section{Model and Circuit equation}
\begin{figure*}[!ht]
	\begin{center}
		\includegraphics[width=0.45\textwidth]{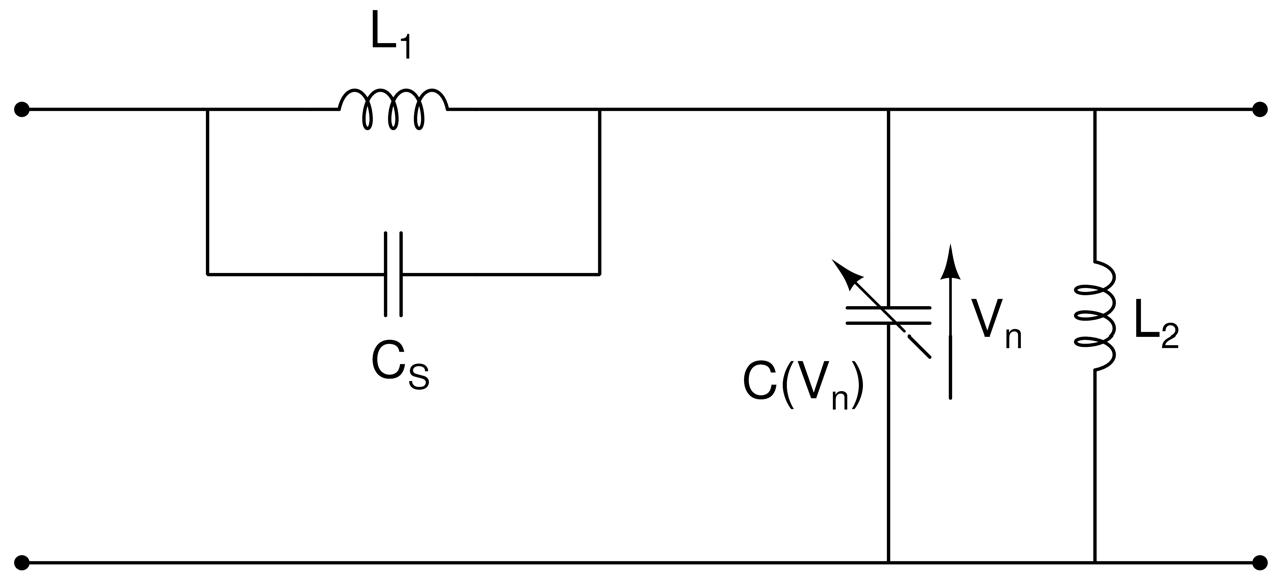}
	\end{center}
    \vspace{0.3cm}
	\caption{Circuit diagram representing a single cell of the discrete nonlinear electrical transmission line}
	\label{fig01}
\end{figure*}

We consider a one-dimensional electrical transmission line with dispersive and nonlinear properties, consisting of $N$ identical cells in sequence. Each cell includes a linear inductor $L_1$ and capacitor $C_S$, facilitating signal transmission and dispersion (see Fig. \ref{fig01}). This setup is connected in series with another parallel configuration of a linear inductor $L_2$ and a nonlinear capacitor $C$, realized via a reverse-biased diode. The nonlinear capacitance $C(V_n + V_b)$ varies with the voltage $V_n$ and follows the polynomial expansion (see Ref. \cite{Kengne2019transmission}):
\begin{equation}
C(V_b+V_n)=\frac{dQ_n}{dV_n}=C_0(1-2\alpha V_n+3\beta V_n^2).
\label{eq1}
\end{equation}
Here, $C_0$ is the characteristic capacitance, while $\alpha$ and $\beta$ are nonlinear coefficients.

Applying Kirchhoff’s laws, we can obtain the wave propagation along the transmission line which is governed by the following discrete differential equation, that is
\begin{align}
\frac{d^2V_n}{dt^2}&+u_0^2(2V_n-V_{n-1}-V_{n+1})-\alpha\frac{d^2V_n^2}{dt^2}+\omega_0^2V_n+\beta\frac{d^2V_n^3}{dt^2}\nonumber\\&+\gamma \frac{d^2(2V_n-V_{n-1}-V_{n+1})}{dt^2}=0,
\label{eq2}
\end{align}
where $u_0=(L_1C_0)^{-\frac{1}{2}}$, $\omega_0=(L_2C_0)^{-\frac{1}{2}}$, and $\gamma=\frac{C_S}{C_0}$ represents dispersion. The first term denotes acceleration at the $n$-th node, while the second term captures coupling between neighboring nodes. The third and fifth terms introduce nonlinear damping or restoring forces, controlled by $\alpha$ and $\beta$, respectively. The fourth term represents a harmonic restoring force with natural frequency $\omega_0^2$. The last term modifies coupling interactions via higher-order derivatives, governed by $\gamma$. Equation (\ref{eq2}) describes a discrete nonlinear system capable of supporting solitons, breather solutions, and rogue waves, depending on parameter values, see for example Refs. \cite{Kengne2017modelling,  Djelah2023first, Kengne2019transmission, Guy2018construction}.

We introduce the slow variables $\zeta$ and $\tau$ as $\zeta=\epsilon(n-v_g t)$ and $\tau=\epsilon^2 t$, where $v_g$ is the group velocity, $n$ is the cell number, and $\epsilon$ is a small parameter. These variables describe the slow evolution of the wave envelope, smoothing rapid variations between cells and enabling a continuous approximation of the system.

We assume the solution of Eq. (\ref{eq2}) in the form \cite{Kengne2019transmission,Djelah2023rogue}:
\begin{equation}
V_n(t)=\epsilon\psi(\zeta,\tau)e^{i\theta}+\epsilon^2\psi_{10}(\zeta,\tau)+\epsilon^2 \psi_{20}(\zeta,\tau)e^{2i\theta}+\text{c.c}.
\label{eq3}
\end{equation}
where $\theta=(kn-\omega t)$ is the rapidly varying phase with wavenumber $k$ and angular frequency $\omega$. The term 'c.c.' represents the complex conjugate. This assumption accounts for the fundamental term $\psi(\zeta,\tau)$, the DC component $\psi_{10}(\zeta,\tau)$, and the second-harmonic term $\psi_{20}(\zeta,\tau)$, which together provide a comprehensive system description.

Substituting Eq. (\ref{eq3}) into Eq. (\ref{eq2}) and retaining terms proportional to $\epsilon$ and $e^{i\theta}$ yield the linear dispersion relation and group velocity:
\begin{equation}
\omega=\sqrt{\frac{\omega_0^2+4u_0^2\sin^2(\frac{k}{2})}{1+4\gamma\sin^2(\frac{k}{2})}}, \quad
v_g=\frac{d\omega}{dk}=\frac{(u_0^2-\gamma\omega^2)\sin(k)}{\omega(1+4u_0^2\sin^2(\frac{k}{2}))}.
\label{eq5}
\end{equation}
Upon solving the resulting equation that arise from $(\epsilon^4, e^{i0})$ provides the expression for the DC term:
\begin{equation}
\psi_{10}(\zeta,\tau)=\frac{2\alpha v_g^2|\psi|^2}{v_g^2-u_0^2}+c_0(\tau)\zeta+c_1(\tau),
\label{eq8}
\end{equation}
where $c_0(\tau)$ and $c_1(\tau)$ are real-valued functions. 

For $(\epsilon^2, e^{2i\theta})$, weobtain the expression for the second-harmonic term as
\begin{equation}
\psi_{20}(\zeta,\tau)=\frac{4\alpha\omega^2\psi^2}{4\omega^2+ 4(4\gamma\omega^2-u_0^2)\sin^2(k)-\omega_0^2}
\label{eq81}.
\end{equation}
Finally, for the case $(\epsilon^3, e^{i\theta})$, we obtain the extended nonlinear Schr\"odinger equation:
\begin{subequations}
	\label{eq9}
\begin{equation}
\label{eq91}
i\frac{\partial \psi}{\partial \tau}+P\frac{\partial^2\psi}{\partial \zeta^2}+Q|\psi|^2\psi+P \beta_1^2  |\psi|^4\psi-4iP\beta_1  (|\psi|^2)_x \psi+\Gamma(\tau)\zeta\psi=0,
\end{equation} 
where
\begin{align}
P&=\frac{1}{1+4\gamma\sin^2(\frac{k}{2})}\Bigg(-\frac{v_g^2}{2\omega}\left(1+4\gamma \sin^2(\frac{k}{2})\right)+\left(\frac{u_0^2}{2\omega}-\frac{\gamma \omega}{2}\right)\cos(k)-2\gamma v_g \sin(k)\Bigg),\\
Q&=\frac{\omega}{1+4\gamma\sin^2(\frac{k}{2})}\Bigg(\frac{3\beta}{2}-\frac{2\alpha^2v_g^2}{v_g^2-u_0^2}-\frac{4\alpha^2\omega^2}{4\omega^2-\omega_0^2+4(4\gamma\omega^2-u_0^2)\sin^2(k)}\Bigg),\\
\Gamma(\tau)&=-\frac{\alpha\omega}{1+4\gamma\sin^2(\frac{k}{2})}c_0(\tau).
\label{eq10}
\end{align}
\end{subequations}
Here, $P$ represents group velocity dispersion, while $Q$ quantifies nonlinear self-modulation.

Equation (\ref{eq91}) can be transformed to the KE equation. To do this, we consider the following transformation,
\begin{subequations}
	\label{eq11}
\begin{equation}
\psi(\zeta,\tau)=\rho \Psi(X,T)e^{i(\Lambda_1 \zeta+\Lambda_2)},
\end{equation}
\text{with}
\begin{align}
X=\Lambda_3+\rho \sqrt{\frac{Q}{2P}}\zeta,\;\; T=\frac{1}{2}Q\rho^2\tau,\;\; \frac{d\Lambda_1}{d\tau}-\Gamma(\tau)=0,\;\; \frac{d\Lambda_2}{d\tau}+P\Lambda_1^2=0,\nonumber\\ \frac{d\Lambda_3}{d\tau}+2P\rho\Lambda_1 \sqrt{\frac{Q}{2P}}=0.
\label{ex1}
\end{align}
\end{subequations}
Substituting the above expressions (\ref{eq11}) into Eq. (\ref{eq9}) and rearranging, we end up at
\begin{equation}
i\frac{\partial \Psi}{\partial T}+\frac{\partial^2 \Psi}{\partial X^2}+2|\Psi|^2\Psi+4\beta_2^2|\Psi|^4\Psi-4i\beta_2 (|\Psi|^2)_x \Psi=0,\quad \beta_2=\rho\beta_1 \sqrt{\frac{2P}{Q}},
\label{eq12}
\end{equation}
the exact KE equation. The KE Eq. (\ref{eq12}) is integrable and it admits several kinds of localized solutions including soliton, breather, RWs and positon.

The breather solution of the KE equation is given by \cite{Solution1}
\begin{subequations}
\begin{equation}
\Psi_{b}(X,T) = \left( \frac{L_{b11}}{L_{b12}} \right) \exp \left( i \left( \rho_0 + \frac{L_{b13}}{L_{b12}} \right) \right)
\label{s1}
\end{equation}
with
\begin{align}
L_{b11} &= -c \eta_1 \cos(\theta_1) - c^2 \cosh(\theta_2) - 2i K_0 \eta_1 \sinh(\theta_2) + 2\eta_1^2 \cosh(\theta_2), \nonumber\\
L_{b12} &= \eta_1 \cos(\theta_1) - c \cosh(\theta_2), L_{b13} = -4\beta_2 \eta_1 K_0 \sin(\theta_1),\nonumber\\
    K_0 &= \sqrt{c^2 - \eta_1^2}, a = -2\xi_1 + 2\beta_2 c^2, b = -a^2 + 4\beta_2^2 c^4 + 2c^2,\nonumber\\ 
    \theta_1 &= 2K_0 (X + 4\xi_1 T), \theta_2 = 4T K_0 \eta_1, \rho_0 = a X + b T.
\end{align}
\end{subequations}
The positon solution for the KE equation is given by \cite{Solution2}
\begin{subequations}
\begin{equation}
    \Psi_{p}(X,T) = \exp \left( -2i\beta_2 \frac{N_{12}(X,T)}{D_1(X,T)} \right) \times \frac{N_{11}(X,T)}{D_1(X,T)}
    \label{s2}
\end{equation}
with 
\begin{align}
    N_{11}(X,T) &= 16 \eta_2 \left( \cos(4\eta_2^2 T - 4T \xi_2^2 - 2X\xi_2) + i \sin(4\eta_2^2 T - 4T\xi_2^2 - 2X\xi_2) \right) \nonumber \\
&\quad \times \left( 8i\eta_2^2 T \cosh(2\eta_2 h_1(X,T)) - 8\eta_2 T \xi_2 \sinh(2\eta_2 h_1(X,T)) \right. \nonumber \\
&\quad \left. -2\eta_2 X \sinh(2\eta_2 h_1(X,T)) + \cosh(2\eta_2 h_1(X,T)) \right)\\
N_{12}(X,T)& = -2 \eta_2 \left( 32 \eta_2 T \xi_2 + 8 \eta_2 X + e^{4\eta_2 h_1(X,T)} - e^{-4\eta_2 h_1(X,T)} \right)\\
D_1(X,T) &= 256 \eta_2^4 T^2 + 256 \eta_2^2 T^2 \xi_2^2 + 128 \eta_2^2 T X \xi_2 \nonumber \\
&\quad + 16 \eta_2^2 X^2 + e^{4\eta_2 h_1(X,T)} + 2 + e^{-4\eta_2 h_1(X,T)}\\
h_1(X,T) &= 4\xi_2 T + X.
\end{align}
\end{subequations}
\begin{figure*}[!ht]
	\begin{center}
		\begin{subfigure}{0.45\textwidth}
			\includegraphics[width=\linewidth]{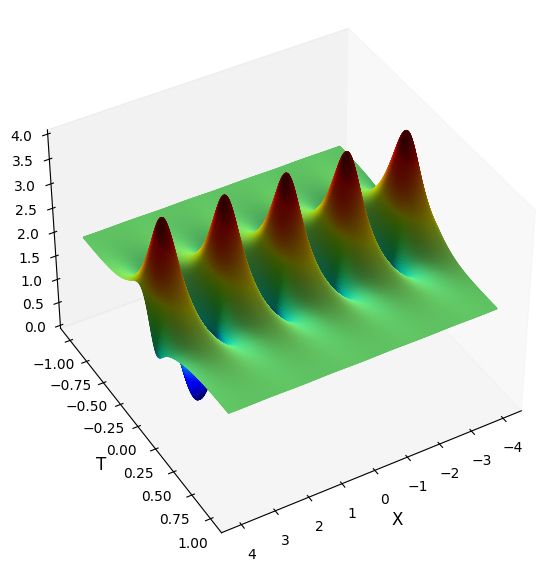}
			\caption{}
		\end{subfigure}
		\begin{subfigure}{0.45\textwidth}
			\includegraphics[width=\linewidth]{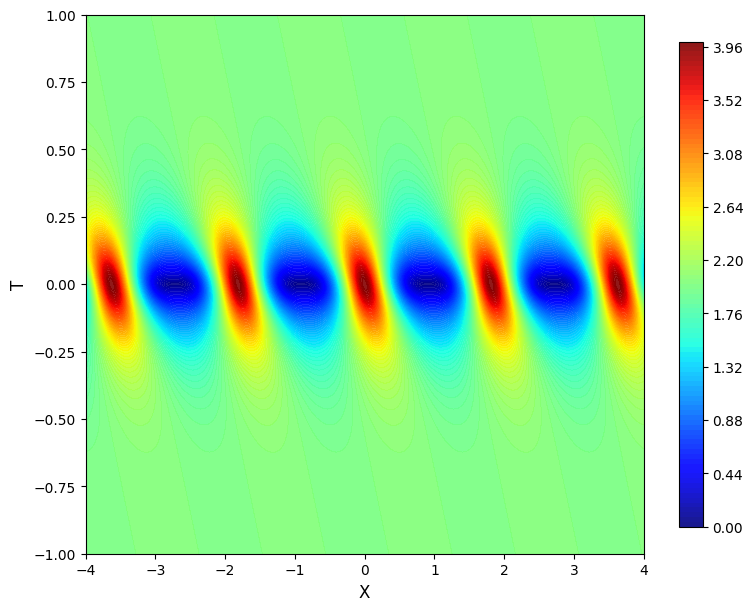}
			\caption{}
		\end{subfigure}\\
		\begin{subfigure}{0.45\textwidth}
			\includegraphics[width=\linewidth]{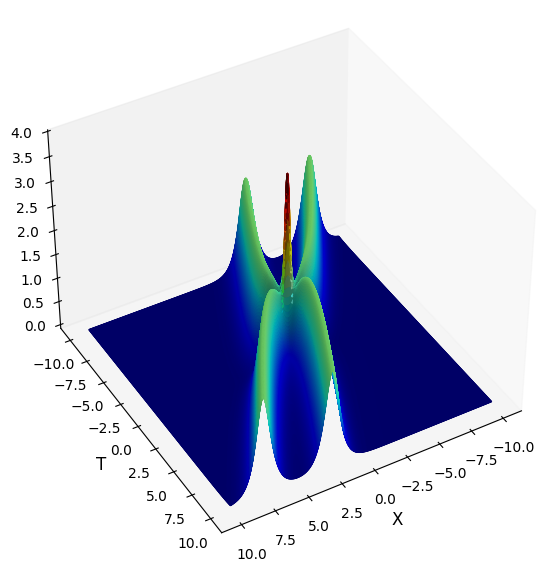}
			\caption{}
		\end{subfigure}
		\begin{subfigure}{0.45\textwidth}
			\includegraphics[width=\linewidth]{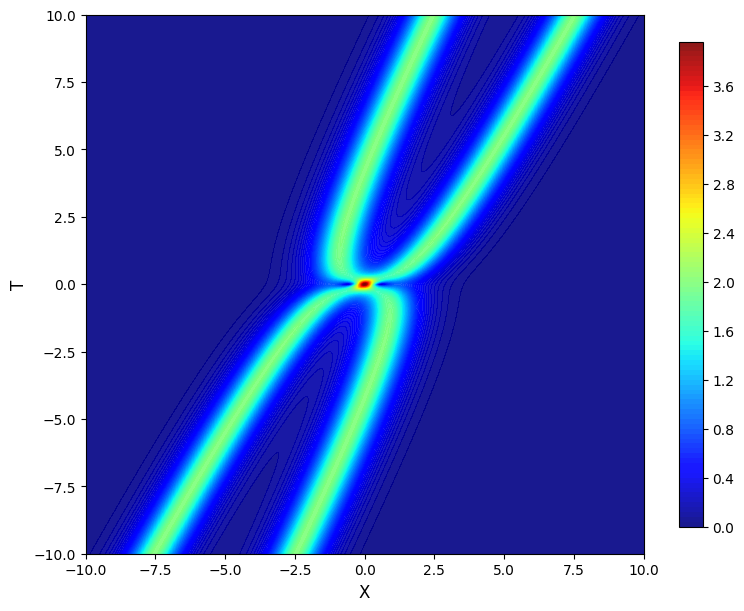}
			\caption{}
		\end{subfigure}
	\end{center}
	\vspace{-0.3cm}
	\caption{Breathers ($|\Psi_{b}(X,T)|$) and positons ($|\Psi_{p}(X,T)|$) solution for the KE Eq. (\ref{eq12}) with parameters: (a) 3D plot for the breather, (b) contour plot for the breather, (c) 3D plot for the positon, and (d) contour plot for the positon.}
	\label{fig10}
\end{figure*} 
Figure \ref{fig10} illustrates the three-dimensional (3D) surface plot and the corresponding contour plot of the breather and positon solutions of the KE equation (\ref{eq12}), depicting their spatiotemporal evolution in the $(X,T)$ plane. Figure \ref{fig10}(a) presents the 3D surface plot of $|\Psi_{b}(X,T)|$, where the breather solution exhibits a localized oscillatory structure concentrated around $T=0$. The amplitude attains its maximum near this region, while for larger values of $|T|$, the oscillations persist but gradually stabilize, leading to an approximately uniform background. The corresponding contour plot in Fig. \ref{fig10}(b) further highlights this behavior, where the color gradient effectively captures the amplitude variations. The breather solution is obtained using the parameter values $c = 2$, $\eta_1 = 1$, $\xi_1 = 0.2$, and $\beta_2 = 0.1$. Similarly, Fig. \ref{fig10}(c) shows the positon solution $|\Psi_{p}(X,T)|$ of the KE equation (\ref{eq12}), while Fig. \ref{fig10}(d) presents its corresponding contour plot. Unlike the breather, the positon solution features a periodic background with a localized, growing wave structure. This solution is obtained for $\xi_2 = -1/8$, $\eta_2 = 1$, and $\beta_2 = 1/4$. In both the breather and positon cases,  we observe that the maximum amplitude reaches approximately $4$.

\section{Dynamical behavior of breather and positon solutions for the KE Eq. (\ref{eq12})}
To study the motion of breathers and positons, we define how their characteristic features evolve over time. For breathers, which consist of multiple oscillations, the motion is best described in terms of the movement of their envelope rather than a single peak. The envelope represents the overall localized region where the oscillations occur. For positons, which are soliton-like structures with slow amplitude modulation, the center refers to the position of the main localized feature that gradually shifts over time. From Eq. (\ref{ex1}), the motion of the positon center or breather envelope is characterized by $\frac{d\zeta}{d\tau}=2P\Lambda_1$ and $\frac{d^2\zeta}{d\tau^2}=2P\Gamma(\tau)$. In the particular case where $\Gamma(\tau) = 0$, the positon remains stationary or undergoes uniform motion depending on whether $\lambda_1(\tau)$ is a zero or non-zero constant. When $\Gamma(\tau) \neq 0$, the positon no longer follows uniform motion; instead, it experiences an acceleration. This means that the trajectory of the positon is influenced by $\Gamma(\tau)$, leading to a variation in its velocity over time. Additionally, the shape of the positon may be affected, leading to modifications in its spatial localization and amplitude distribution. For breathers, $\Gamma(\tau)$ plays a crucial role in determining their oscillatory behavior. When $\Gamma(\tau) = 0$, the breather exhibits periodic oscillations in space-time, and its envelope propagates uniformly if $\lambda_1(\tau)$ is constant. However, when $\Gamma(\tau) \neq 0$, the breather's oscillation frequency and wavelength become time-dependent, leading to a dynamic modulation of its profile. This can result in a shift of the breather's phase, a change in its periodicity, or even a deformation in its wave structure, depending on the specific form of $\Gamma(\tau)$. For this reason, in our work, we have considered $\Gamma(\tau)$ as (i) a constant, (ii) periodic form and (iii) exponential form. 
\subsection{Case (i): Constant linear potential $\Gamma(\tau)=A_0$}
Here, we consider a time-dependent linear potential, meaning $\Gamma(\tau) = A_0$, where $A_0 \neq 0$. Substituting this into the $X$-dependent relation in Eq. $(\ref{ex1})$ yields $X$ in the following form:
\begin{equation}  
    X = \rho \sqrt{\frac{Q}{2P}} \left[ \zeta - \frac{P(A_0 \tau + d_1)^2}{A_0} \right] + d_2,
\label{x1}  
\end{equation}  
where $d_1$ and $d_2$ are real constants. Using Eq. $(\ref{x1})$ along with the breather and positon solutions (Eqs. $(\ref{s1})$ and $(\ref{s2})$) for the KE equation, we can analyze the dynamics of breathers and positons under a time-dependent linear potential.
\begin{figure*}[!ht]
	\begin{center}
		\begin{subfigure}{0.45\textwidth}
			\includegraphics[width=\linewidth]{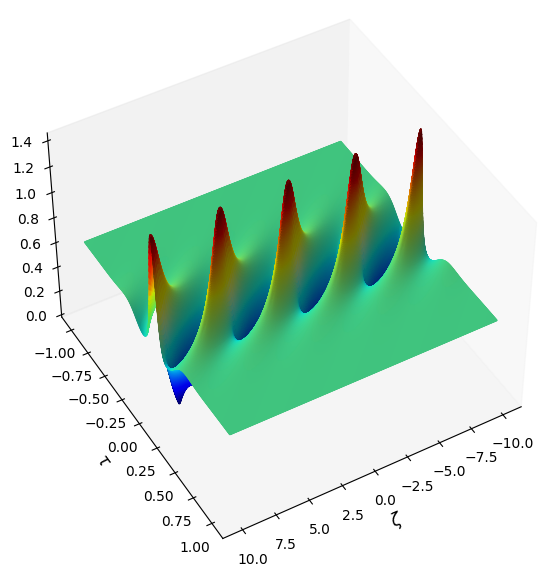}
			\caption{}
		\end{subfigure}
		\begin{subfigure}{0.45\textwidth}
			\includegraphics[width=\linewidth]{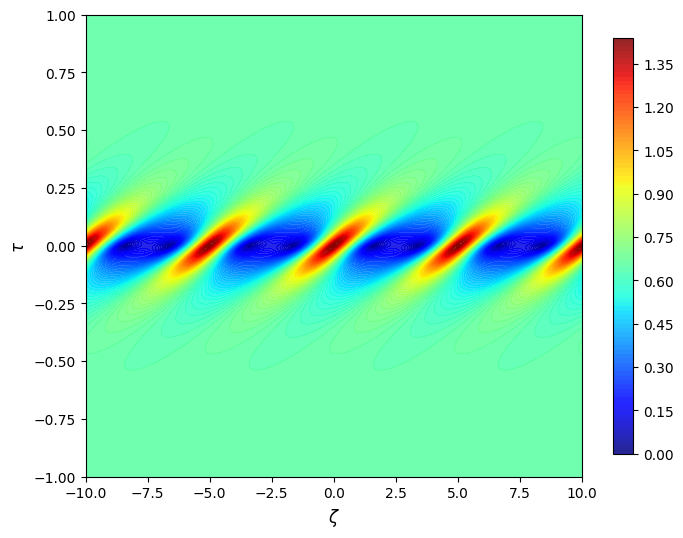}
			\caption{}
		\end{subfigure}\\
		\begin{subfigure}{0.45\textwidth}
			\includegraphics[width=\linewidth]{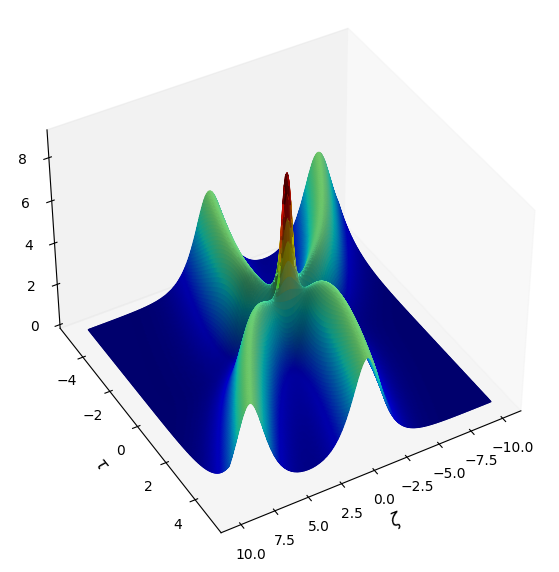}
			\caption{}
		\end{subfigure}
		\begin{subfigure}{0.45\textwidth}
			\includegraphics[width=\linewidth]{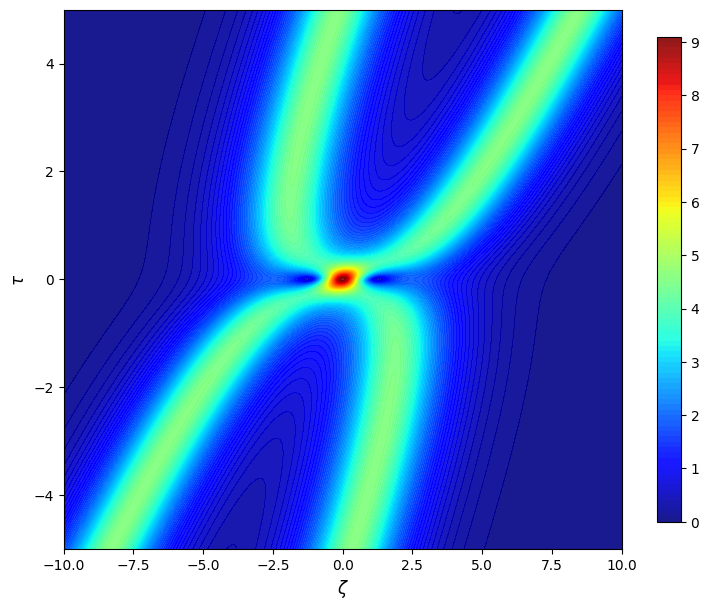}
			\caption{}
		\end{subfigure}
	\end{center}
	\vspace{-0.3cm}
	\caption{Breathers ($|\Psi_{b}(\zeta,\tau)|$) and positons ($|\Psi_{p}(\zeta,\tau)|$) in the electrical transmission line (\ref{eq2}) with constant potential and $A_0 = 0.001$, $d_1 = 0$,  $d_2 = 0$: (a) 3D plot for the breather, (b) contour plot for the breather, (c) 3D plot for the positon, and (d) contour plot for the positon.}
	\label{figb1}
\end{figure*} 

For the following figures, we generally fix the parameters as $L_1 = 220 \times 10^{-3}$ H, $L_2 = 470 \times 10^{-3}$ H, $C_0 = 370 \times 10^{-6}$ F, and $C_S = 240.5 \times 10^{-6}$ F, with $c_{01} = 0$, $c_1 = 0$, $c_2 = 0$, and $k = 0.75$. For breather solutions, we consider $ \alpha = -0.5 $, $ \beta = -0.1445 $, $ \rho_0 = 1.3 $, $ \beta_2 = 0.1 $, $ c = 0.5 $, $ \eta_1 = 0.3 $, and $ \xi_1 = 0.2 $. For positon solutions, the parameters are $ \alpha = 0.21 $, $ \beta = 0.0197 $, $ \rho_0 = 2.3 $, $ \xi_2 = -\frac{1}{8} $, $ \eta_2 = 1 $, and $ \beta_2 = \frac{1}{4} $.

Figure $\ref{figb1}$ presents the propagation of breather and positon solutions in the nonlinear electrical transmission line modeled using a modified Naguchi circuit with a linear time-dependent constant potential. The breather solution is computed over the spatial and temporal range $ \zeta \in [-10,10] $ and $ \tau \in [-1,1] $, whereas the positon solution is computed over $ \zeta \in [-10,10] $ and $ \tau \in [-5,5] $. Figures $\ref{figb1}(a)$ and $\ref{figb1}(c)$ depict the three-dimensional surface plots of $ |\psi(\zeta,\tau)| $, while Figures $\ref{figb1}(b)$ and $\ref{figb1}(d)$ show the corresponding contour plots. The maximum amplitude of the breather is approximately $1.4$, whereas for the positon solution, it is $8$. From the results, we observe that the orientation of the breather changes compared to Figures $\ref{fig10}(a)$ and $\ref{fig10}(b)$. Additionally, when a constant potential is applied, the maximum amplitude of the breather decreases. Similarly, the positon solutions also exhibit a change in orientation; however, in contrast, their amplitude increases under a constant time-dependent linear potential. This suggests that the applied potential plays a crucial role in modulating both the shape and amplitude of the breather and positon waves, highlighting its influence on wave propagation in the nonlinear transmission line.

\subsection{Case (ii): Temporal periodic modulation of the function of the linear potential}
Now, we consider a periodic time-dependent linear potential, given by $\Gamma(\tau) = \frac{1}{2P} \left[ 2 + 45\omega_1^2 \cos(\omega_1 \tau) \right]$, where $\omega_1$ is a real constant. Substituting this into the $X$-dependent relation in Eq. $(\ref{ex1})$, we obtain:
\begin{equation}
    X = \rho \sqrt{\frac{Q}{2P}} \left[ \zeta - 2d_3 P \tau - \tau^2 + 45 \cos(\omega_1 \tau) \right] + d_4,
\label{x2}
\end{equation}
where $d_3$ and $d_4$ are real constants. Using Eq. $(\ref{x2})$ along with the breather and positon solutions (Eqs. $(\ref{s1})$ and $(\ref{s2})$) for the KE equation, we can analyze the impact of a temporally periodic linear potential on the dynamics of breathers and positons.
\begin{figure*}[!ht]
	\begin{center}
		\begin{subfigure}{0.45\textwidth}
			\includegraphics[width=\linewidth]{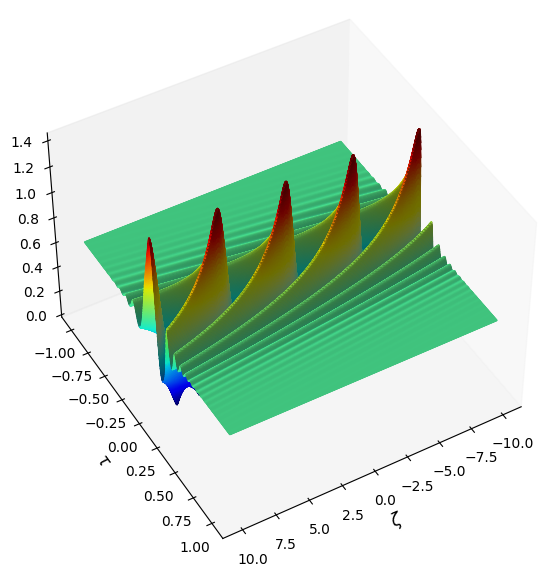}
			\caption{}
		\end{subfigure}
		\begin{subfigure}{0.45\textwidth}
			\includegraphics[width=\linewidth]{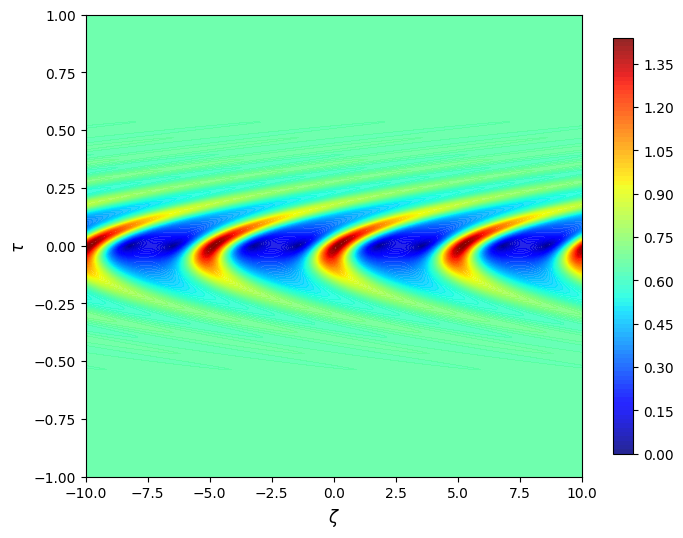}
			\caption{}
		\end{subfigure}\\
		\begin{subfigure}{0.45\textwidth}
			\includegraphics[width=\linewidth]{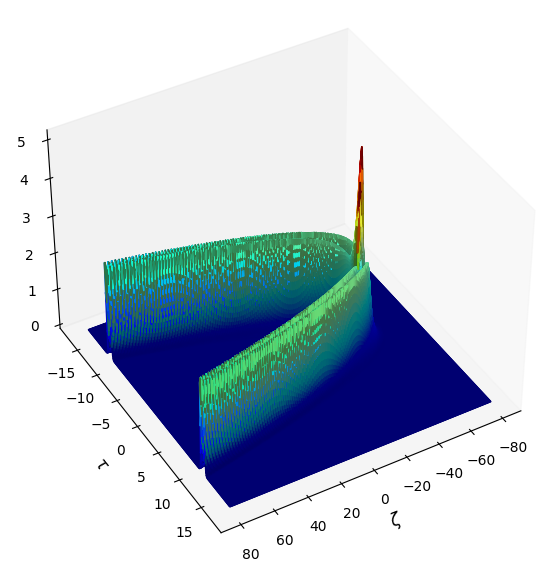}
			\caption{}
		\end{subfigure}
		\begin{subfigure}{0.45\textwidth}
			\includegraphics[width=\linewidth]{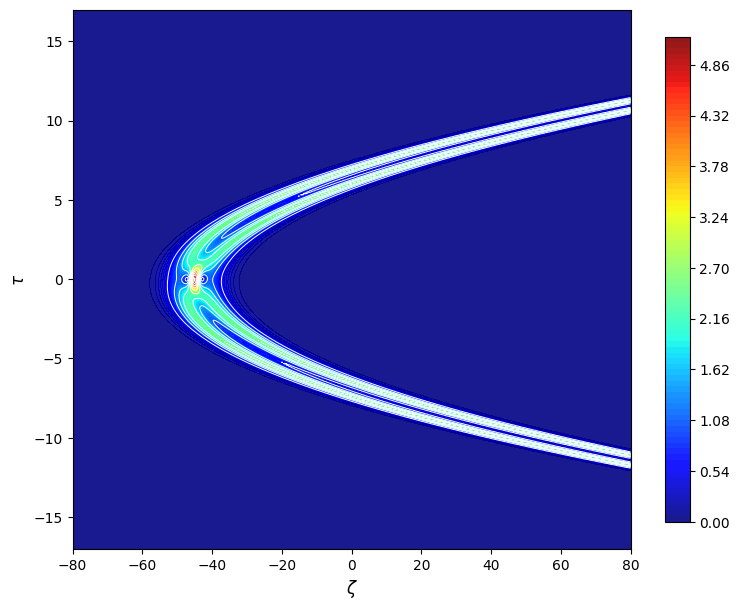}
			\caption{}
		\end{subfigure}
	\end{center}
	\vspace{-0.3cm}
	\caption{Breathers ($\omega_1=2$) and positons ($\omega_1=0.0001$) in the electrical transmission line (\ref{eq2}) with periodic potential and : (a) 3D plot for the breather ($|\Psi_{b}(\zeta,\tau)|$), (b) contour plot for the breather, (c) 3D plot for the positon ($|\Psi_{p}(\zeta,\tau)|$), and (d) contour plot for the positon.}
	\label{fig3dpp}
\end{figure*} 

The plots presented in Fig. $\ref{fig3dpp}$ analyzes the breather and positon solutions in a nonlinear electrical transmission line modeled using a modified Naguchi circuit with a linear temporal periodic potential. The first column (Figs. $\ref{fig3dpp}(a)$ and $\ref{fig3dpp}(c)$) shows the three-dimensional surface plots of $ \psi(\zeta,\tau) $, while the second column (Figs. $\ref{fig3dpp}(b)$ and $\ref{fig3dpp}(d)$) presents the corresponding contour plots. Here, also we set the integration constants $ d_3 = 0 $ and $ d_4 = 0 $. The parameter $ \omega_1 = 2 $ is used for the breather solution, whereas $ \omega_1 = 0.0001 $ is used for the positon solution. The remaining parameters are the same as those given above. The solution is computed over the spatial and temporal range $ \zeta \in [-10,10] $ and $ \tau \in [-1,1] $ for the breather, while for the positon, it is computed over $ \zeta \in [-80,80] $ and $ \tau \in [-17,17] $. The maximum amplitude of both waves occurs at different locations compared to the previous two cases (Figs. $\ref{fig10}$ and $\ref{figb1}$). From the results, we observe that the periodic potential (temporal periodic) significantly influences the wave structures. The breather solution transforms into a crescent-shaped breather, exhibiting curved localization patterns distinct from standard breathers. Similarly, the positon solution forms a crescent-shaped positon, characterized by a curved and elongated structure. These findings suggest that the applied periodic potential modulates both the shape and orientation of the waves, leading to unique waveforms in the nonlinear transmission line.
\subsection{Case (iii): Temporal exponential modulation of the function of the linear potential}
Finally, we consider an exponentially varying time-dependent linear potential, given by $\Gamma(\tau) = b_0 e^{b_1 \tau}$, where $b_0$ and $b_1$ are nonzero real constants. Substituting this into the $X$-dependent relation in Eq. $(\ref{ex1})$, we obtain:
\begin{equation}
    X = \rho\sqrt{\frac{Q}{2P}} \left[ \zeta - 2P \frac{b_0}{b_1^2} e^{b_1 \tau} - 2P d_5 \tau \right] + d_6
\label{x3}
\end{equation}
where $d_5$ and $d_6$ are real constants. Using Eq. $(\ref{x3})$ along with the breather and positon solutions (Eqs. $(\ref{s1})$ and $(\ref{s2})$) for the KE equation, we can analyze the influence of an exponentially varying linear potential on the dynamics of breathers and positons.
\begin{figure*}[!ht]
	\begin{center}
		\begin{subfigure}{0.45\textwidth}
			\includegraphics[width=\linewidth]{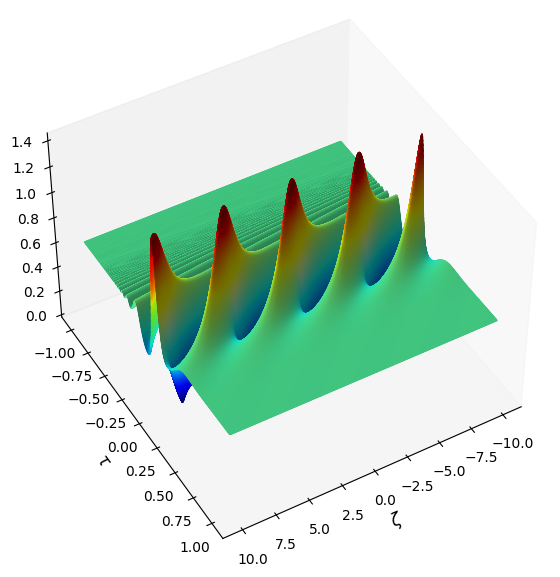}
			\caption{}
		\end{subfigure}
		\begin{subfigure}{0.45\textwidth}
			\includegraphics[width=\linewidth]{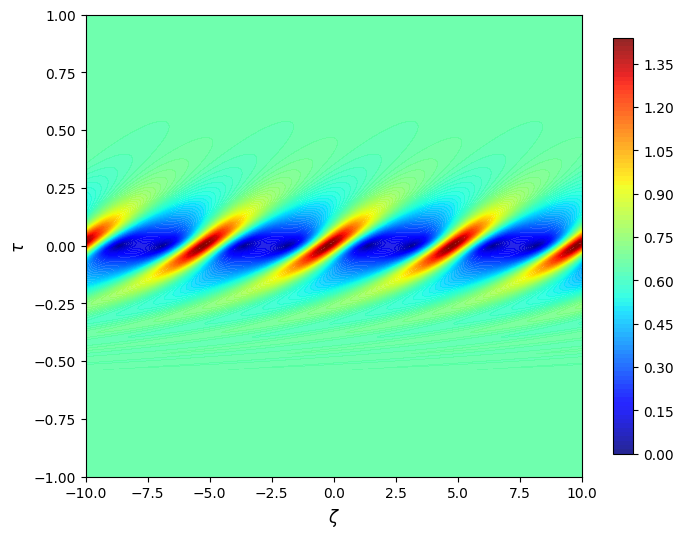}
			\caption{}
		\end{subfigure}\\
		\begin{subfigure}{0.45\textwidth}
			\includegraphics[width=\linewidth]{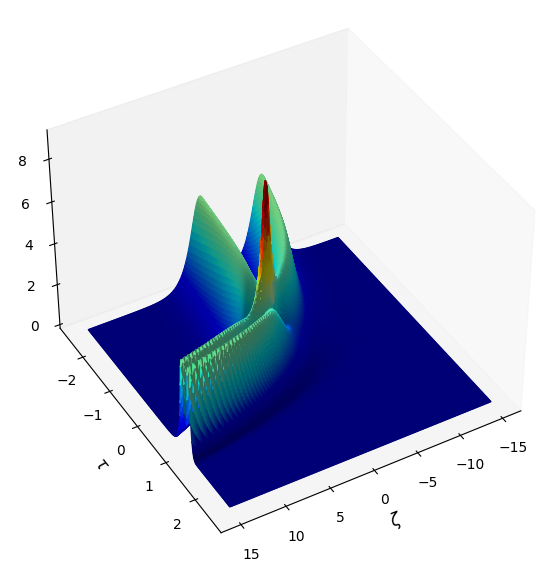}
			\caption{}
		\end{subfigure}
		\begin{subfigure}{0.45\textwidth}
			\includegraphics[width=\linewidth]{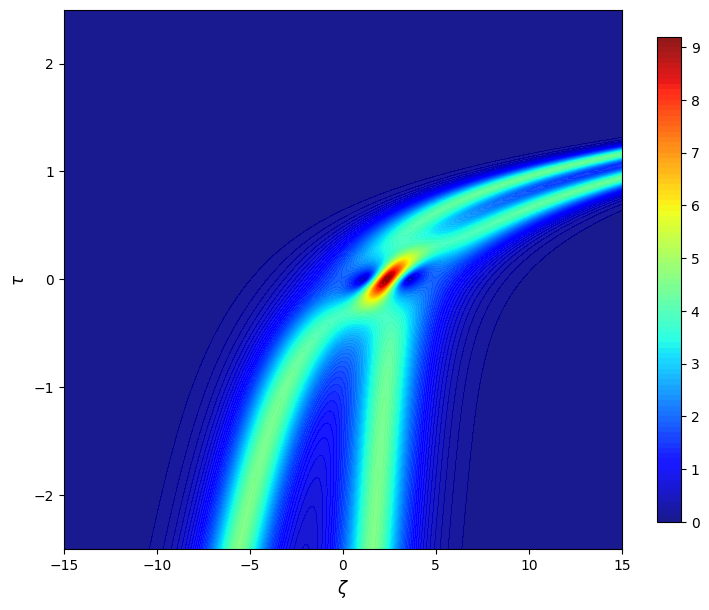}
			\caption{}
		\end{subfigure}
	\end{center}
	\vspace{-0.3cm}
	\caption{Breathers($|\Psi_{b}(\zeta,\tau)|$)  and positons ($|\Psi_{p}(\zeta,\tau)|$) in the electrical transmission line (\ref{eq2}) with exponential potential: (a) 3D plot for the breather, (b) contour plot for the breather, (c) 3D plot for the positon, and (d) contour plot for the positon.}
	\label{fig3dep}
\end{figure*} 

Figure $\ref{fig3dep}$ presents the analysis of breather and positon solutions in a nonlinear electrical transmission line modeled using a modified Naguchi circuit with a linear time-dependent constant potential. Figures $\ref{fig3dep}(a)$ and $\ref{fig3dep}(c)$ denote the three-dimensional plots, while Figs. $\ref{fig3dep}(b)$ and $\ref{fig3dep}(d)$ show the corresponding contour plots. Here, we consider a nonzero constant value of $ b_0 = 1 $, while the integration constants $ d_5 = 0 $ and $ d_6 = 0 $. For the parameter $ b_1 $, we set $ b_1 = P $ for breather solutions and $ b_1 = P/2 $ for positon solutions. The remaining parameters are the same as in the previous cases. The solution is computed over the spatial and temporal ranges $ \zeta \in [-10,10] $ and $ \tau \in [-1,1] $ for breathers, and $ \zeta \in [-15,15] $ and $ \tau \in [-2.5,2.5] $ for positons. The maximum amplitude attained is approximately $ 1.4 $ for the breather and $ 8 $ for the positon. The results indicate that the exponential modulation potential significantly affects the wave structures. The breather solution evolves into an asymmetric crescent-shaped breather, displaying a curved localization with an uneven intensity distribution. Similarly, the positon solution forms an asymmetric crescent-shaped positon, characterized by a curved and elongated profile with an imbalanced structure. These findings highlight the role of the exponential modulation potential in altering both the shape and orientation of the waves, leading to distinct asymmetric formations in the nonlinear transmission line.
\section{Breathers and positons in electrical transmission line} In the previous section, we analyzed the breather and positon solutions with experimental coordinates. Now, we analyze both solutions in the electrical transmission line context. To do this, we consider the second-order positon solution of the circuit equation (\ref{eq2}), which is expressed in the form given in Eq. (\ref{eq3}) and is given by:
\begin{equation}
V_n(t)=Re(\epsilon\psi_{b/p}(\zeta,\tau)e^{i\theta}+\epsilon^2\psi_{10}(\zeta,\tau)+\epsilon^2 \psi_{20}(\zeta,\tau)e^{2i\theta}+c.c.),
\label{ep31}
\end{equation}
where $\psi_{b/p}(\zeta,\tau)$, $\psi_{10}(\zeta,\tau)$ and $\psi_{20}(\zeta,\tau)$ are given in Eqs. (\ref{s1}) or (\ref{s2}), (\ref{eq8}) and (\ref{eq81}) respectively. The term $ \psi_{b/p} $ denotes that when considering $ \psi_b $, the breather solution is substituted, and when considering $ \psi_p $, the positon solution is used.
\begin{figure*}[!ht]
	\begin{center}
		\begin{subfigure}{0.45\textwidth}
			\includegraphics[width=\linewidth]{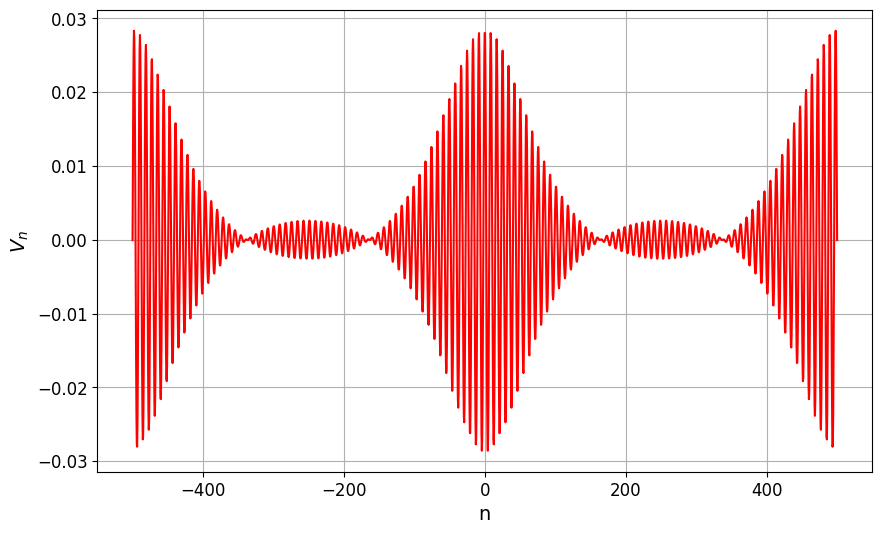}
			\caption{}
		\end{subfigure}
		\begin{subfigure}{0.45\textwidth}
			\includegraphics[width=\linewidth]{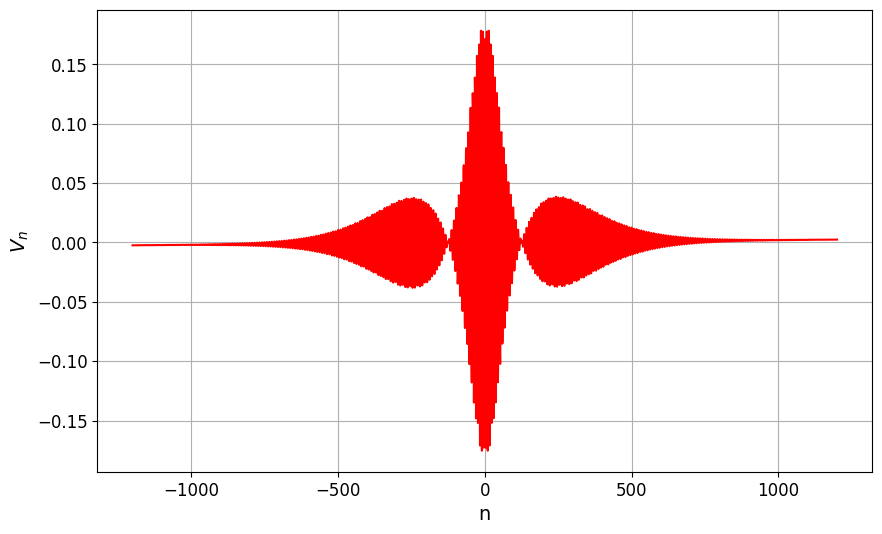}
			\caption{}
		\end{subfigure}\\
		\begin{subfigure}{0.45\textwidth}
			\includegraphics[width=\linewidth]{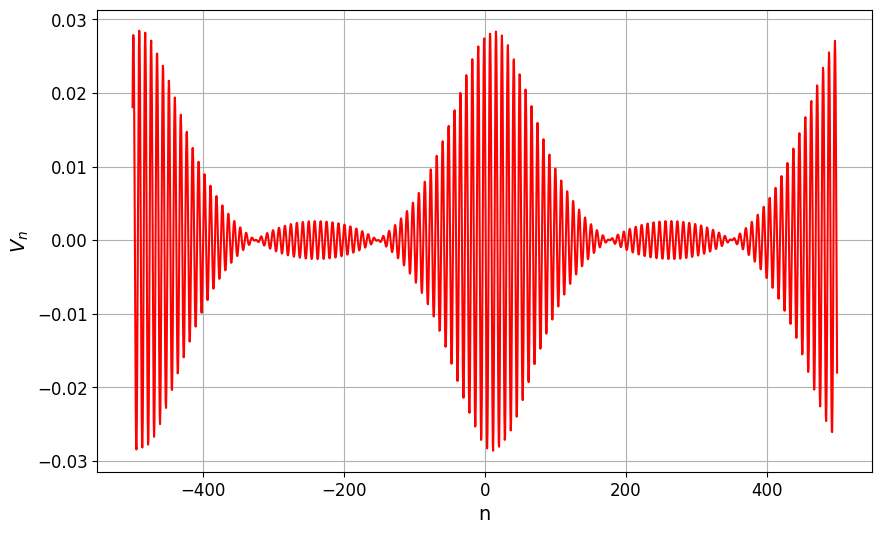}
			\caption{}
		\end{subfigure}
		\begin{subfigure}{0.45\textwidth}
			\includegraphics[width=\linewidth]{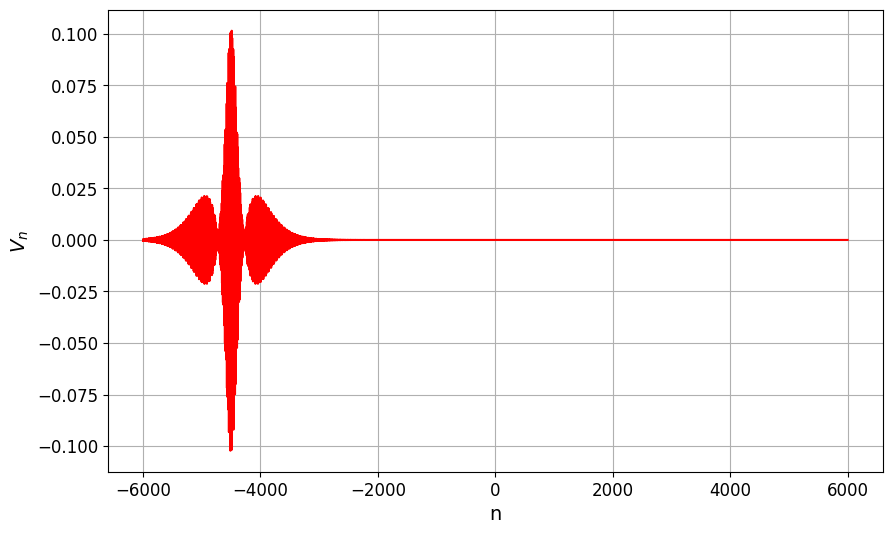}
			\caption{}
		\end{subfigure}\\
		\begin{subfigure}{0.45\textwidth}
			\includegraphics[width=\linewidth]{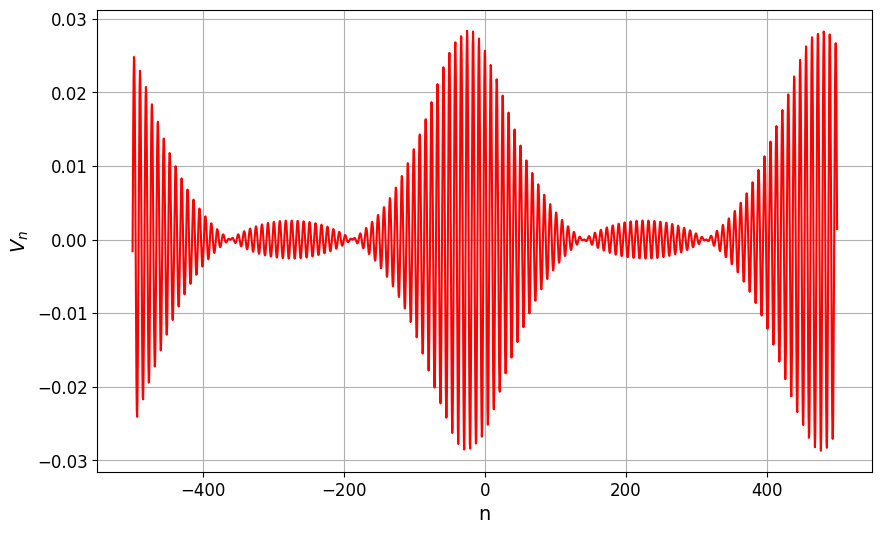}
			\caption{}
		\end{subfigure}
		\begin{subfigure}{0.45\textwidth}
			\includegraphics[width=\linewidth]{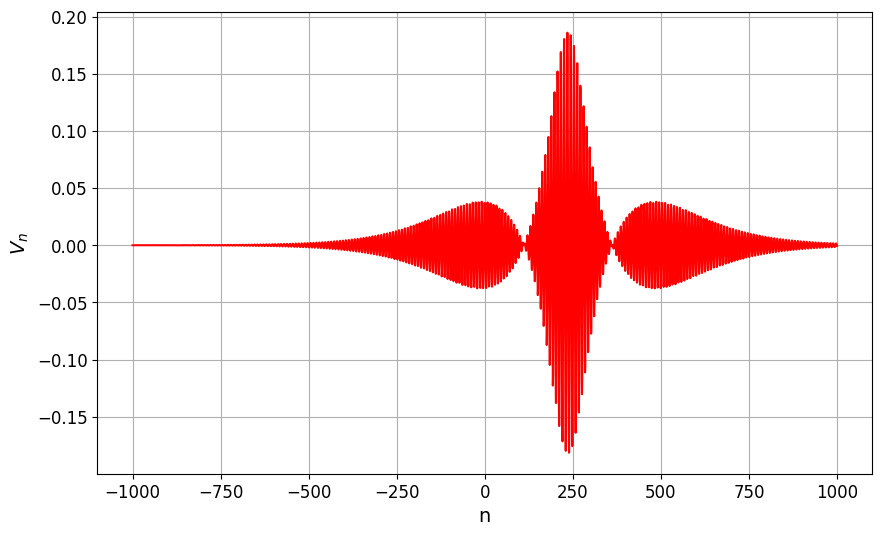}
			\caption{}
		\end{subfigure}
	\end{center}
	\vspace{-0.3cm}
	\caption{Breathers and positons in the electrical transmission line (\ref{eq2}) using (\ref{eq3}) and $\epsilon=0.01$: (a) Constant potential for breather, (b) Constant potential for positon,  (c) temporal periodic potential for breather, (d) temporal periodic potential for positon, (e) exponential potential for breather and (f) exponential potential for positon.}
	\label{figbp}
\end{figure*} 

In Fig. \ref{figbp}, we demonstrate the dynamical behavior of breathers and positons in the modified Noguchi electrical transmission line. Here, we investigate the solution dynamics with three types of linear potentials:  (i) constant time-dependent, (ii) temporally periodic, and (iii) exponential at $t=0$. With these linear potentials, the voltage $V_n$ is plotted. The first column (Figs. \ref{figbp}(a), \ref{figbp}(c), \ref{figbp}(e)) represents breather solutions under constant, periodic, and exponential potentials, respectively. The second column (Figs. \ref{figbp}(b), \ref{figbp}(d), \ref{figbp}(f)) represents positon solutions under the same potentials. For the constant potential, the central peaks of the breather attain their maximum level at $n = 0$ within the range $-400 \leq n \leq 400$. Under a periodic cosine potential, the maximum amplitude of the breather peaks shifts forward, meaning the localization of breather peaks changes within the same $n$-region. When we apply an exponential potential, the maximum amplitude of the central breather peaks moves backward, shifting into the negative $n$-region. Similarly, for the positon solution, the range of $n$ is $-1200 \leq n \leq 1200$ for the constant potential, $-6000 \leq n \leq 6000$ for the periodic potential, and $-1000 \leq n \leq 1000$ for the exponential potential. As in the case of breathers, for the constant potential, the maximum amplitude of the positon appears at $n = 0$. Under a periodic potential, the positon peaks shift backward into the negative $n$-region. Compared to breathers, positons are more sensitive, as their backward shift is more pronounced. Under an exponential potential, the positon moves forward, meaning it shifts into the positive $n$-region.
\section{Conclusion}
In this study, we investigated breathers and positons in an electrical transmission line modeled by the modified Naguchi electrical transmission line using the KE equation. By employing the reductive perturbation method and a specific transformation, we analyzed the behavior of these wave structures. Furthermore, we examined the effects of different time-dependent linear potentials on the dynamics of breathers and positons in the KE equation. Three distinct cases were considered:

\begin{itemize}
    \item Constant linear potential $\Gamma(\tau) = A_0$: This potential significantly modifies the characteristics of breather and positon solutions. The orientation of both wave structures is noticeably altered, indicating an influence on their spatial evolution. Additionally, the breather solution experiences a reduction in amplitude, suggesting a suppression of its localized oscillatory behavior. In contrast, the positon solution exhibits an increase in amplitude while maintaining its shape, implying an enhancement in wave intensity. These findings emphasize the critical role of external potentials in modulating wave dynamics within nonlinear transmission lines.
    \item Periodically modulated potential  $\Gamma(\tau) = \frac{1}{2P} \left[ 2 + 45\omega_1^2 \cos(\omega_1 \tau) \right]$:  Under this modulation, the breather solution transforms into a crescent-shaped breather with curved localization, while the positon solution evolves into a crescent-shaped positon with an elongated profile. This result demonstrates that periodic potentials not only influence wave orientation but also introduce distinct structural modifications, thereby affecting wave propagation in the nonlinear transmission line.

    \item Exponentially varying potential $\Gamma(\tau) = b_0 e^{b_1 \tau}$:  This type of modulation plays a crucial role in shaping wave dynamics, leading to the formation of asymmetric crescent-shaped breather and positon structures. It affects their orientation, intensity distribution, and overall profile, highlighting the significant impact of time-dependent potentials on nonlinear wave propagation.
\end{itemize}
The above results present the breather and positon solutions for Eq. (\ref{eq91}). Additionally, we studied the solution dynamics in the transmission line using the relation $V_n$. The results show that linear potentials significantly impact breather and positon dynamics in the modified Noguchi transmission line. Constant potentials keep peaks at $n=0$, periodic potentials shift breathers forward and positons backward, while exponential potentials move breathers backward and positons forward.

Previous studies have primarily focused on the effects of such external modulations on rogue waves. In contrast, our work explores breathers and positons, offering new insights into their behavior under different time-dependent linear potentials. Additionally, we have studied their effects in the electrical transmission line using the relation $V_n$, which was not considered in previous studies on rogue waves  \cite{Djelah2023rogue}. Our results underscore the importance of external modulation in controlling wave dynamics, with potential applications in nonlinear optical systems, Bose-Einstein condensates, and nonlinear transmission lines. 

Furthermore, we observe that the maximum amplitude of the positon solution is notably higher for the KE equation compared to the NLS equation, with clear differences also seen in the orientation of the waves. Similar amplitude behavior is observed in the electrical transmission line model, supporting the theoretical predictions. For breather solutions, a similar amplitude trend is present, although their localization differs between systems. These comparative insights strengthen our previous findings and highlight the unique dynamical features of the KE equation. These changes arise due to the influence of higher-order nonlinear terms and related effects.

In future, we have plane to explore the role of additional external potentials in coupled systems to deepen the understanding of their impact on nonlinear wave evolution.
\section*{Acknowledgments}
 NS wishes to thank DST-SERB, Government of India for providing National Post-Doctoral Fellowship under Grant No. PDF/2023/000619. The work of MS forms a part of a research project sponsored by Council of Scientific and Industrial Research (CSIR) under the Grant No. 03/1482/2023/EMR-II. K.M. acknowledges support from the DST-FIST Programme (Grant No. SR/FST/PSI-200/2015(C)).

\end{document}